\documentclass[sigconf,screen]{acmart}

\setcopyright{acmcopyright}
\copyrightyear{2024}
\acmYear{2023}
\acmDOI{} % \acmDOI{XXXXXXX.XXXXXXX}
\acmConference[ Unpublished ]{N/A}{Sep.}{2024}
\acmISBN{N/A} %\acmISBN{XXX-X-XXXX-XXXX-X/XX/XX}

\def\Description#1{\relax}
\def\lili{\textsc{Lili}}
\def\fsharp{\textsc{F\#}}
\def\ensico{\textsc{Ensico}}
\def\python{\textsc{Python}}
\def\jupyter{\textsc{Jupyter}}
\def\haskell{\textsc{Haskell}}

\begin{document}

\title{First Steps towards K-12 Computer Science Education in Portugal --- Experience Report}%{K-12 CS Education in Portugal}
%% If title is too long for the header, please provide a shorter title (must fit on one line, no breaks!):
%% \title[Shorter Title]{Long Title}

%author{\ensico\ Association}
%email{info@ensico.pt}
%authornotemark[1]
%
\author{F. L. Neves}
\affiliation{%
  \institution{\ensico, %streetaddress{
       Largo Dr.\ Tito Fontes, 15}
  \city{Porto}
  \country{Portugal}
  %\postcode{4000-538}
}
\email{f.luis.neves@ensico.pt}
\authornote{Both authors contributed equally to this research.}
\author{J. N. Oliveira}
\affiliation{%
  \institution{\textsc{INESC TEC}, Univ. Minho, \ensico}
  %\streetaddress{Campus de Gualtar}
  \city{Braga}
  \country{Portugal}}
\email{jno@di.uminho.pt}
\orcid{0000-0002-0196-4229}
%

%% If the list of authors is too long for the header, please provide a more
%% concise list:

%% The abstract is a short summary of the work to be presented in the paper.
\begin{abstract}

Computer scientists Jeannette Wing and Simon Peyton Jones have catalyzed a pivotal discussion on the need to introduce computing in K-12 mandatory education. In Wing's own words, computing \emph{represents a universally applicable attitude and skill set everyone, not just computer scientists, would be eager to learn and use.}

The crux of this educational endeavor lies in its execution. This paper reports on the efforts of the \ensico\ association to implement such aims in Portugal. Starting with pilot projects in a few schools in 2020, it is currently working with 4500 students, 35 schools and 100 school teachers. The main aim is to gain enough experience and knowledge to eventually define a comprehensive syllabus for teaching computing as a mandatory subject throughout the basic and secondary levels of the Portuguese educational system.

A structured framework for integrating computational thinking into K-12 education is proposed, with a particular emphasis on mathematical modeling and the functional programming paradigm. This approach is chosen for its potential to promote analytical and problem-solving skills of computational thinking aligned with the core background on maths and science.

%  A clear and well-documented \LaTeX\ document is presented as an article formatted for publication by ACM in a conference proceedings or journal publication. Based on the ``acmart'' document class, this article presents and explains many of the common variations, as well as many of the formatting elements an author may use in the preparation of the documentation of their work.

\end{abstract}

%%
%% The CCSXML block below contains important meta data for the digital
%% library. Please generate adequate meta data for your paper at
%% http://dl.acm.org/ccs.cfm and replace the whole CCSXML block.
%% Also, the tool generates the CCS concepts part of the paper, thus also
%% the ccsdesc commands below the CCSXML block should be replaced accordingly.
%%




\maketitle

\section{Introduction}

Computer scientists Jeannette M. Wing
%\footnote{https://www.linkedin.com/in/jeannette-wing-1b88a63/}
(USA) and Simon Peyton Jones
%\footnote{https://www.linkedin.com/in/simonpj/}
(UK) have catalyzed a pivotal discussion on the need to introduce a K-12 computing syllabus into mandatory education \cite{Wi06,kolling2013bringing}. This trend was galvanized by Wing's influential paper, '\emph{Computational Thinking}' (CT), in which she posits that computing '\emph{[...] represents a universally applicable attitude and skill set everyone, not just computer scientists, would be eager to learn and use.}' However, the crux of this educational endeavor lies not in its justification but in its execution, and the literature does not yet show {consensus} on the broader inclusion of Computer Science (CS) in mandatory K-12 education \cite{DBLP:journals/eait/WebbDBKRCS17,Xu2023,VF20}. Quoting \cite{DBLP:journals/jcal/SunHZ21}:
\begin{quote}\em
    [...] however, there is no unified conclusion on how to design programming activities to promote the acquisition of CT skills more effectively.
\end{quote}

\href{https://ensico.pt/}{\ensico} \cite{Ensico} is a private, non-profit association founded in Portugal (2019) to address the challenge mentioned above that believes that incorporating computing into the first twelve years of education can promote equity and enhance scientific and technological literacy. Moreover, this has the potential to foster creativity and multidisciplinary learning, as well as to improve oral and written communication skills, understanding of formal concepts, and awareness of scientific history.
As a member of the \href{https://www.informaticsforall.org/}{\textsc{Informatics for All}} coalition, \ensico\ follows the \href{https://www.informaticsforall.org/the-informatics-reference-framework-for-school-release-february-2022/}{Informatics Reference Framework} \cite{I4A}, which advocates that

\begin{quote}
    [...] \em informatics should exist as a discipline at all stages of the school curriculum, starting early in primary school and continuing to exist and develop through upper secondary school. Moreover, we suggest that education in informatics should be compulsory for all pupils from primary through secondary education, having a status and standing similar to that of language and mathematics. Well-educated teachers and teacher-teachers are essential to realise this vision.
\end{quote}

 Starting with pilot projects in a few schools, \ensico\ is currently working with 4500 students, 35 schools and 100 school teachers. Its main aim is to gain enough experience and knowledge to eventually define a comprehensive program for teaching computing as a mandatory subject throughout the basic and secondary levels of the Portuguese educational system. However, this raises some  questions: should the focus be on technology or science, programming or general computing, concepts or procedures, formative or informative education, training or entertainment?

This paper aims to address these pivotal questions and propose a structured framework for integrating computational thinking into K-12 education, with a particular emphasis on mathematical modeling and the functional programming paradigm. This approach is chosen for its potential to promote the analytical and problem-solving skills that are central to computational thinking, naturally aligned with the core background on maths and science.

% \color{red} .... The first has to do with how the K-12 educational system is structured in each case. The second is dependent on the first --- the earlier you start, the less technical you should be, avoiding a too early dependence on technology. The third is perhaps the most challenging: what should be actually taught, and in what order? Technology or science? Just computer programming or computing in general? Concepts or recipes? Formative or informative? Training or entertaining?

Although the bulk part of \ensico's activity takes place in Portugal, an international arm was developed in parallel between 2020 and 2023 through the ERASMUS+ project \textsc{CS4All} \cite{CS4ALL}, in partnership with The Open University, Cisco ASC in the UK and Colectic in Spain. % This initiative aims to share the successful outcomes and learning from the ongoing project in Portugal, which is focused on formal education. Additionally, it seeks to enhance these experiences with innovative methodologies and approaches. These improvements are being developed in collaboration with our partners: These collaborations are particularly focused on the realms of informal and non-formal education.

The remainder of this paper is structured as follows: section \ref{sec:basic}
briefly addresses basic principles concerning the overall teaching plan.
Details about \ensico's efforts to frame itss plan into the Portuguese K-12 educational
system are given in section \ref{sec:method}. The efforts since 2020 to fulfill
this plan are presented in section \ref{sec:plan}. Section \ref{sec:contents}
provides details about classes and their different teaching styles. The organization
of events and \ensico's presence on social media are the subject of section
\ref{sec:media}. Finally, section \ref{sec:eval} concludes and gives an outlook
for future activities.

\section{Basic principles}\label{sec:basic}

There are three basic challenges when adding a new topic to an educational track \cite{DBLP:journals/eait/WebbDBKRCS17}: \textsc{when} to start, \textsc{what} to teach and \textsc{how} to teach. Concerning \textsc{when} and \textsc{how}, the \ensico\ general principle is that initiation to computing can start at early ages, and it does not require much equipment: just a piece of paper, a pencil and, above all, grey matter. Essentially, a new problem is thrown into the classroom: how to \emph{communicate with a machine}? And how different is this from communicating with people?

Where communicating with people requires written and spoken language proficiency, communicating with machines requires what is nowadays called \emph{computational thinking} \cite{Wi06}. This requires logical thinking and the ability to abstract from complex situations in real life into simple models that computers understand.

\begin{figure}%[h]
  \centering
  \includegraphics[width=0.75\linewidth]{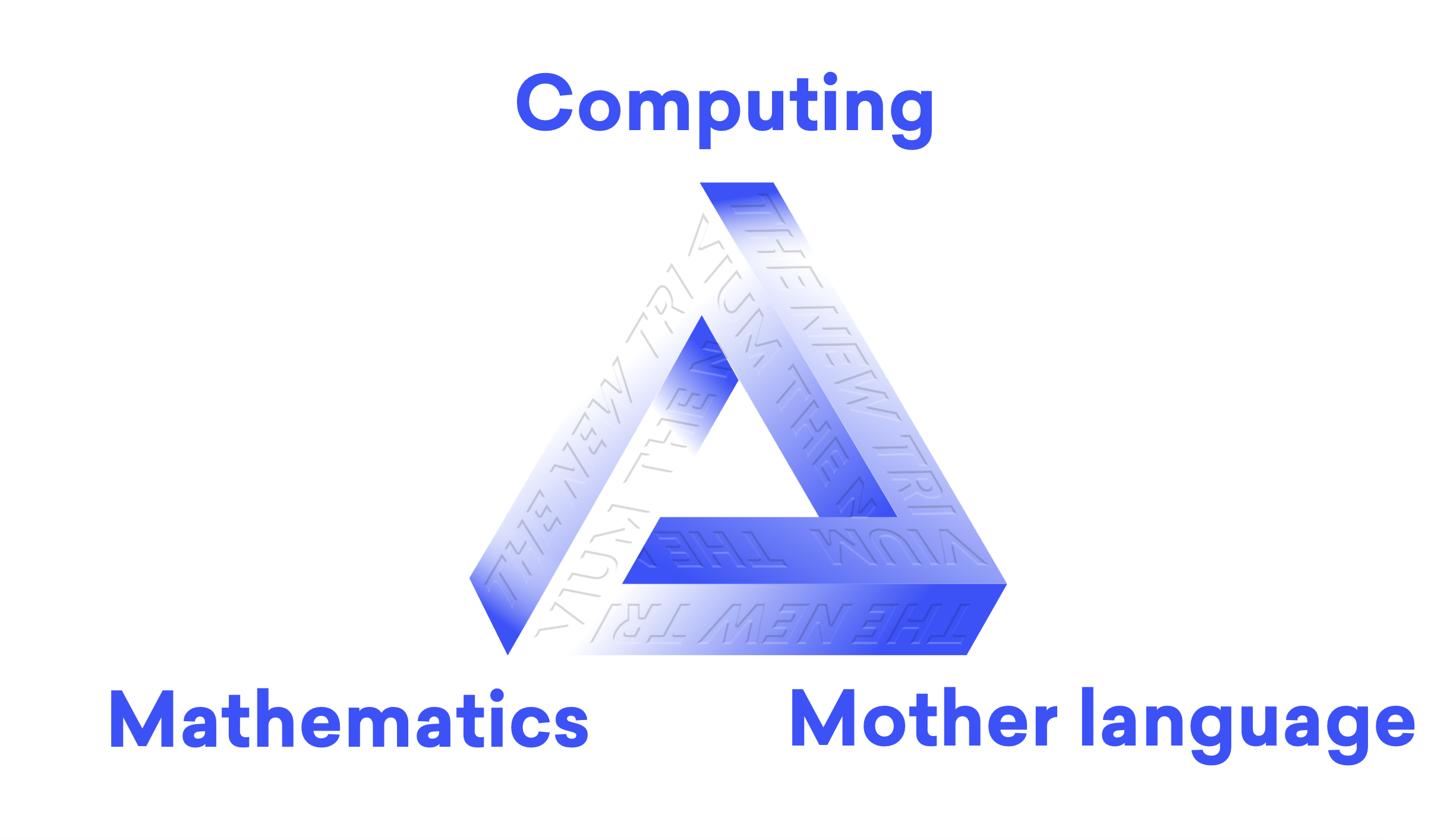}
  \caption{Introducing computing in K-12 education calls for a ``New Trivium'' \cite{Bu03} able to heal the ``big divide'' between science and the humanities, while linking computing to its very foundations on maths \cite{Kr07}.\label{fig:tnt}}
  \Description{TNT}
\end{figure}

This triggers immediate synergies with two other core subjects: \emph{natural language}, since no one can convey to a computer what cannot be communicated to another person; and \emph{mathematics}, since this is the form of expression that better deals with abstract information that computers are able to understand (Fig.~\ref{fig:tnt}).

Concerning the \textsc{what} challenge above, one honestly must admit that it traverses CS education as a whole, in particular at the university level.\footnote{See e.g.\ the recommendations by the ACM/IEEE-CS Task Force on Computing Curricula \cite{ACM01}. }
In his landmark book "Algorithms + Data Structures = Programs"
%, published in 1976 by Prentice-Hall
\cite{Wi76}, Niklaus Wirth --- the creator of the programming language PASCAL
who sadly passed away in January of this year (2024) --- 
shows a constant pedagogical concern starting from a basic intuition that
programming theory later confirmed:
\begin{quote}
[...] \em data precede algorithms: you must have some objects before you can perform operations on them.
\end{quote}
(quote from the Preface). Strangely, many introductory programming courses still ignore, almost half a century later, such a relevant message --- a message of which the book provides ample evidence \cite{Wi76}.

\ensico\ follows this intuition, which is all the more relevant the younger the students are: even before data, one should start with the physical things that computer data (will later) model, i.e.\ \emph{dematerialize}. Computing terminology is highly metaphoric precisely because of such a need to abstract from physical entities --- think e.g.\ of terms such as ``stack'', ``queue'', ``tree", ``memory'' and so on.
Clearly, children need to understand such physical entities prior to abstracting from them. This matches nicely with the ``unplugged'' method \cite{DBLP:journals/cacm/Bell21} that is followed for the very early K-12 stages, as explained later concerning \textsc{how} to teach.\footnote{One cannot \emph{dematerialize} what one does not materially know about. In a sense, \ensico\ lets children start from the ``metaphors [they will later] program by" \cite{LJ80}.}

Two basic data structures shine in  such a \emph{data-oriented} introduction to computing --- lists and pairs. It is surprising how so many concepts \cite{Jac21} can be taught on such a simple basis\footnote{Just a few examples: bit maps, look-up tables, key-value stores, midi (music), 2D-graphics, matrices, finite state automata, dictionaries... --- the list is long!} as the pedagogical materials already produced show (section \ref{sec:contents}).

Data processing, i.e \emph{transformations}, comes next. Children literature is full of fantastic tales in which things \emph{transform} into other things by \emph{magic}. Now, what else can better capture the transformation of thing $x$ into thing $y$ by ``magic'' $f$ than the proper concept of a mathematical function, $y=f(x)$? May this serve as a simple explanation for the decision to follow the functional paradigm \cite{BW88} when it comes to move from the \emph{unplugged} to the \emph{plugged} phases of the teaching sequence.

A thorough explanation of this and other pedagogical decisions is, however, out of scope of the current paper, which is intended to focus on the difficult thing --- the path that is being followed concerning the \emph{execution challenge} mentioned above.

\section{The Context}\label{sec:method}

The Portuguese K-12 educational system comprises two levels: basic and secondary school. Children in Portugal usually begin basic school at age 6, which spans nine years and is segmented into a first cycle (ages 6 to 9), a second cycle ages 10 to 11) and a third cycle (12 to 14). Students typically reach secondary school at age 15, which continues for three years.

\subsection{Basic School}

\begin{figure}%[h]
  %\centering
  \includegraphics[width=0.3\textwidth]{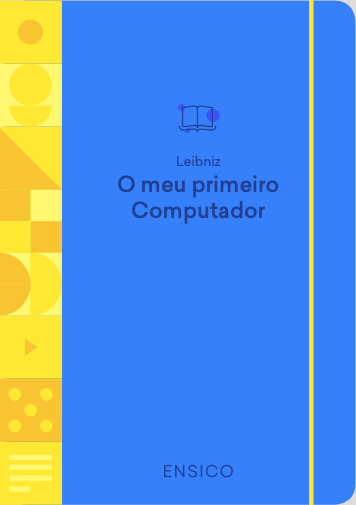}
   \caption{The  '\textsc{Leibniz}' notebook, \ensico\ students' "first computer", promotes the \emph{CS-unplugged} teaching method pioneered by Tim Bell \cite{DBLP:journals/cacm/Bell21}.\label{fig:leibniz}}
  \Description{leibniz.}
\end{figure}

\begin{figure}%[h]
  %\centering
  \includegraphics[width=1.0\linewidth]{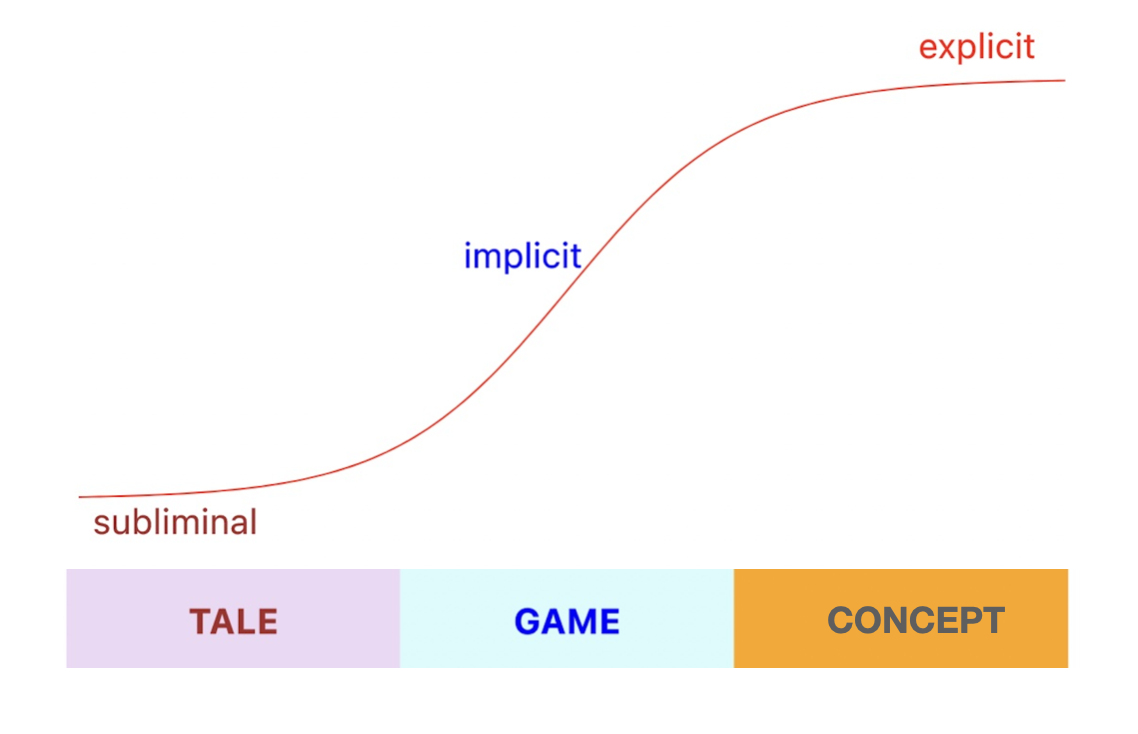}
   \caption{\ensico's knowledge ``modulation wave": starting from \emph{subliminal} messages hidden in little stories or tales (first years), knowledge becomes \emph{implicit} in games or the physical manipulation of concrete objects (middle years) before becoming explicit in \emph{concepts} \cite{Jac21} and, finally, \emph{programs} in the ``plugged" phase.\label{fig:wave}}
  \Description{wave.}
\end{figure}

In the first two years of the first cycle (ages 6--7), \ensico's methodology focuses on learning without the use of computers (Fig.~\ref{fig:leibniz}). The learning starts through stories with captivating characters that implicitly link to computing subjects (Fig.~\ref{fig:lili}). These characters create emotional connections with the students that allow for the exploration of computational themes in a funny and pedagogically effective way. That is, knowledge begins to be acquired subliminally and indirectly, becoming progressively more explicit and direct as years advance (Fig.~\ref{fig:wave}). In the final two years of the first cycle (ages 8--9), the aim is to leverage the teaching of computing in the learning of Mathematics and Portuguese (and potentially foreign languages).

The introductory computer classes start in a \emph{semi-plugged} style, \jupyter\ Notebook-based exercises directed by the teacher, followed by \emph{plugged}, hands-on sessions conducted by the students themselves using \jupyter\ Notebooks. At this stage, the exercises consist basically of evaluation simple expressions in the purely functional \haskell\ programming language \cite{DBLP:journals/jfp/Jones03g}. Haskell was chosen due to the simplicity of its syntax, bearing very little \emph{language impedance} when compared to the maths expressions that students find in other textbooks.\footnote{The IHaskell kernel \cite{Gi07} is used to enable Haskell interpretation in Jupyter cells. Haskell's minimalist syntax is enabled by its sophisticated type inference, which allows concise code without explicit type annotations --- an advantage recognized by practitioners \cite{MarketSplash24}.}

The second cycle (ages 10--11) further integrates computing with the study of mathematics and languages. It delves into linguistic concepts essential for mastering the native language and mathematical terminology, laying the groundwork for programming. Building on their initial computer experiences from the first cycle, students progress to creating their first data models and algorithms.
% employing the purely functional \haskell\ programming language.

The third cycle (ages 12--14) aims to consolidate learning going deeper into the \emph{function} concept alongside the math curriculum, which begins the study of functions at this stage. Consequently, the functional programming paradigm is deliberately embraced. The frequency of semi-plugged and plugged classes increases, eventually leading up to the beginning of collaborative programming activities and the adoption of multi-paradigm programming languages such as \fsharp\ and Python.

\subsection{Secondary School}

Secondary school (ages 15--17) building upon the foundations of basic education, \ensico's plans at this level are to emphasize the study and exploration of computationally significant fields, including Cybersecurity, Artificial Intelligence, Big Data, the Internet of Things, Blockchain, and more. This exploration is facilitated by advanced programming environments and languages endowed with extensive and educationally valuable libraries, such as \python. At this stage, one should aim to expand and develop domain-oriented specialised modules in collaboration with various universities, which will be available to students as elective courses tailored to their personal and professional interests.

\section{Action Plan}\label{sec:plan}

Central to the overall strategy is \ensico's roll-out of its five-year pilot program, segmented into six phases. The initial phase (2020), that included only school teachers, was followed by five phases that incorporate both students and teachers from pilot schools. These participate in weekly computing classes over five consecutive school years, spanning from 2020/21 to 2024/25. Since day one, \ensico's strategy has embraced an experimental, bottom-up approach, focusing initially on engaging school teachers, directors, students and parents. For this initiative \ensico\ has obtained significant financial support, predominantly from private entities and local authorities (see Section \ref{sec:ackn}). Moreover, frequent dialogue established with the Portuguese presidency, the government, municipalities and educational stakeholders aims to ensure the necessary commitment for a final political decision on the nationwide implementation of \ensico's computing program.

\subsection{March to July 2020, Phase I}

The pilot project started with a sequence of webinars, 18 hours in total, involving 25 school teachers from both basic and secondary school and coming from 3 different schools, two public and one private, located in Oporto.

The involvement of teachers of a large spectrum of disciplines, such as math, ICT, physics, natural sciences, Portuguese, English and music, among others, was crucial to check their awareness of the opportunity to incorporate such a body of knowledge in the Portuguese curriculum. It was equally important at this stage to clarify concepts and present \emph{computer science} (CS) as a potential mandatory topic, and showing how well it can blend with other courses, in particular with Portuguese and Mathematics. Questionnaires filled at the end revealed a positive perception about the inclusion of CS into the Portuguese curricula:

\begin{itemize}
\item 82\% of teachers expressed interest in teaching CS concepts within their discipline;
\item 55\% of teachers expressed interest in becoming CS teachers;
\item 100\% of teachers believed that CS had the potential to help students perform better in the subjects they taught;
\item 91\% of teachers considered that acquiring CS skills would substantially contribute to students' future;
\item 90\% of teachers identified possible synergies between CS and their own teaching discipline;
\item The majority of teachers considered CS skills to be equally or slightly more important than skills in Mathematics or English;
\item 75\% of teachers found it important to make CS education mandatory in the first three cycles of learning, while 27\% considered it important for secondary school as well.
\end{itemize}

\begin{figure}%[h]
  %\centering
  \includegraphics[width=0.47\textwidth]{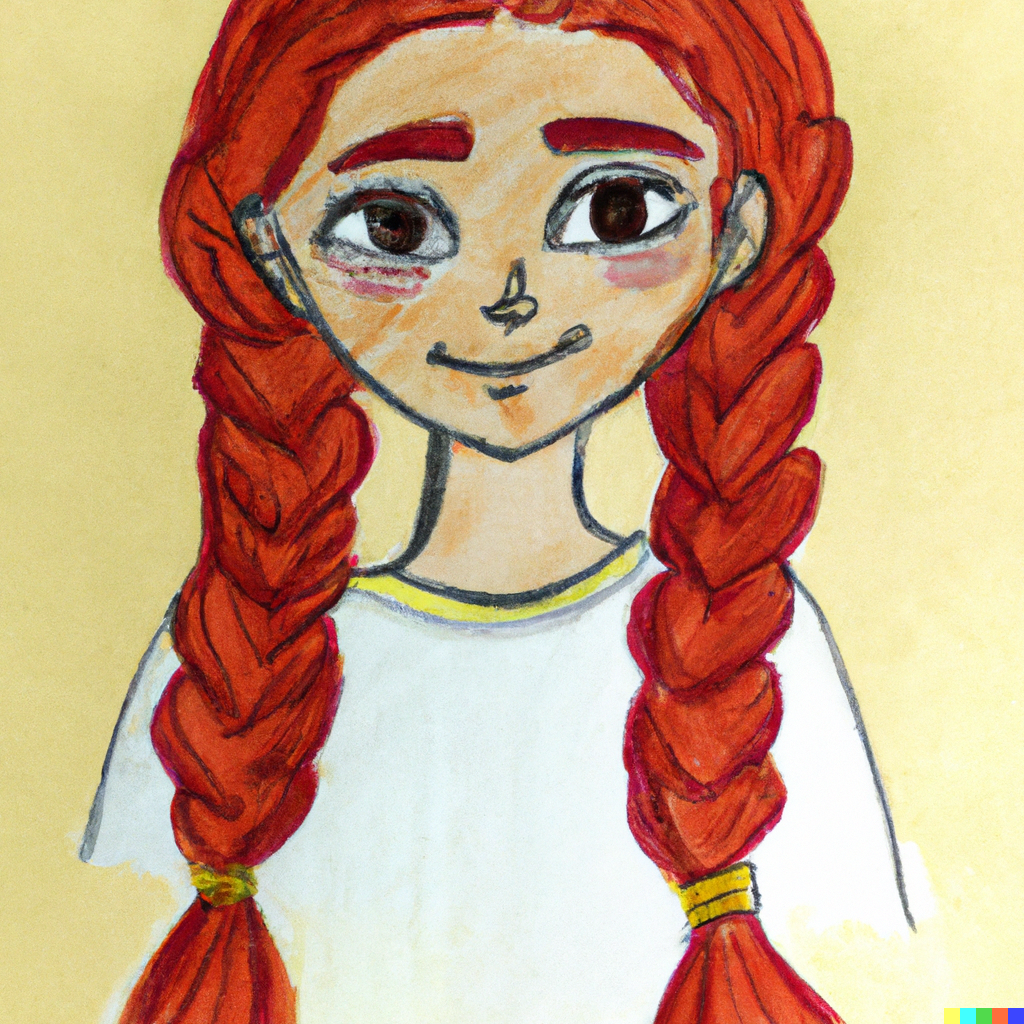}
   \caption{Born tongeless, \lili\ needs to communicate using a binary, non-verbal language. Lili's adventures introduce children to the binary language and binary communication codes, initiating a path that will lead them to the Braille and Morse codes, to bitmaps (Fig.~\ref{fig:bmap}), QR-codes (Fig.~\ref{fig:qr}) and so on and so forth. \label{fig:lili}}
  \Description{lili.}
\end{figure}

\subsection{July 2020 to June 2024, Phases II to V}

From July 2020 onwards the different pilot phases established the step-by-step execution plan that had started in just 3 schools and 8 classes of grades 5 to 8.

The main focus of the current plan for CS education lies in students in the second and third cycles. As stressed by Kramer \cite{Kr07}, it is on this age range that, according to Jean Piaget, the cognitive transition between the 'concrete' stage (ages 6--11) to the 'formal' or 'abstract' stage (ages 12--17) takes place. Computer education
%, as advocated by \ensico,
should explore this crucial period of cognitive development.
% to apply different methods and significant transitions throughout the 12 years of learning.

The main goal of the first year of the pilot project was the production of pedagogical material and its evaluation in the classroom, involving students and teachers. The joint work with school teachers was of a major importance to promote their CS and digital skills while conducting continuous training on computing contents at the same time.

Between July 2021 to June 2023 there was an expansion of the teaching plan, reaching grade nine. Also important was the geographic expansion, not in terms of quantity, but in terms of representative schools/groups across Portugal. \ensico's goal is to create a program as inclusive as possible, reflecting both regional and gender realities and sensibilities. Running mainly in the Oporto and Lisbon regions, the pilot is currently spreading to schools in the Douro Valley, Minho and Trás-os-Montes provinces. % and other places in the north and center of the country.

In the 2021/22 school year, a total of 1216 computing lessons, each lasting 50 minutes and covering grades 3 to 9, were conducted. These sessions were spread over 40 classes, with each class receiving around 32 lessons throughout the year.

% The final grades were mostly above 65\%. {Grades, per year (average)68\% 75\%  65\%  77\%  59\%  66\%  68\% Grade 3   Grade 4  Grade 5  Grade 6  Grade 7  Grade 8  Grade 9}

% At the end of each period (3 per year), assessments were conducted for the different classes, school teachers and parents/guardians. It was found that 63\% of first cycle students enjoyed all the subjects taught. Among these, ASCII, Morse codes and the representation of color images were the most appreciated. Furthermore, in this age group, 93\% of the students considered computer classes to be useful for them, especially because they found the subjects very interesting and enjoy learning more about computers.

% It was also found that 44\% of students in the second and third cycles enjoyed all the subjects taught in the first year of computing. Among these, ASCII, Morse codes and the binary representation of images were the most favored. Additionally, 79\% of these students considered the computer science subjects to be useful.

% Similarly, it was observed that 33\% of students in the second and third cycles liked all the subjects taught in the second year of computing. Among these, functional programming and computer graphics were the most appreciated. Furthermore, 83\% of these students found the computer science subjects to be useful.

%Still in relation to the students and in global terms:

\def\omitted{
\begin{itemize}
\item On a scale of 0 to 10, an average score of 8 was given by students regarding how likely they would recommend computer classes to a friend, family member, or acquaintance;
\item The favorite activities of 40\% of the students involved pencil and paper;
\item 49\% of the students would like to have more online classes, especially to watch more videos about the subjects and to clarify doubts with the teacher.
\end{itemize}
}

% With respect to school teachers, it was found that 92\% of the surveyed teachers were interested in learning more about computing / computational thinking. Furthermore, 92\% of the surveyed teachers saw possible synergies between computing and the subject they teach. Finally, 54\% of the teachers were of the opinion that computer classes should be mandatory for all levels of basic education, and 54\% of the surveyed teachers believe that this discipline should have a mandatory aspect even in secondary education.

% Last but not least, with respect to parents/guardians, it was found that 70\% of them believe that computer education should be mandatory. Among these, 62\% advocate for computing education to be compulsory in all years of basic and secondary education. Finally, 76\% of the parents/guardians who responded to the survey would like their children to attend a school that includes computer classes in its educational plan.

The current and the next school years (2023--2025) are regarded as vital for the national expansion of the \ensico's computing program. For the first time, a formal training program started to prepare (mainly first cycle and maths) teachers to lecture K-12 computing classes to their students. As shown in table \ref{tab:pilot}, 100 teachers were trained in 2023 to start lecturing computing classes in the 2023/24 school year. In 2024, we expect to train more than 200 teachers, reaching more than 300 in total, in order to expand the computing classes in the 2024/25 school year. These "new computing teachers" are trained to follow the proposed computing program and to use the provided materials, which include slides, activities, \jupyter\ Notebooks, etc.

The first group of 100 teachers are being coordinated and evaluated since September 2023 and the results so far seem promising. The majority of schools in  the municipality of Cascais have adopted \ensico's computing education program.
For the next four years, \ensico\ will conduct an impact assessment in collaboration with the \href{https://www.novasbe.unl.pt/en/}{Nova School of Business and Economics} (see Section \ref{sec:eval}).

\begin{table}
  \caption{Pilot Evolution\label{tab:pilot}}
  %\label{tab:pilot}
  \begin{tabular}{rcccc}
%   \toprule
    \textbf{Number of} & 2020/21 & 2021/22 & 2022/23 & 2023/24\\
\hline %    \midrule
    students & 172 & 1600 & 2500 & 4500\\
    teachers &   8 &   62 &  100 &  300\\
    schools & 4 & 21 & 28 & 35\\
    classes & 280 & 2555 & 3990 & 7945\\
% \bottomrule
\end{tabular}
\end{table}

\section{About the Classes}\label{sec:contents}

% The story begins with a mute female character. Her name consists of two letters: \lili\ (Fig. \ref{fig:lili}). She devises a binary language to communicate with others and interface with a computer.

% Then, other characters appear along the way, such as textbf{flip} (the \textbf{bit flip} function) and \textbf{keep} (the \textbf{identity} function). Suddenly, they begin to transform bit maps and, with such power, they change a cat into a dog, a dog into a heart, a closed padlock into a key, and a key into an open padlock that unlocks the grand door of the castle. With arrows and more arrows, computing students will register these magic transformations upon  matrices populated by \textbf{flip} and \textbf{keep} functions.

This section briefly explains how the basic principles of section \ref{sec:basic}
get implemented by \ensico\ in the classroom. The \emph{golden principle} of always conveying
new knowledge on top of knowledge students already have (Jean Piaget) calls
for \emph{backwards chaining} the cognitive stages of Fig.~\ref{fig:wave}. That
is, if content $C$ is to be taught at year $Y_n$, then there should be some
content $C'$ taught at year $Y_m$ ($m < n$) such that $C'$ prepares $C$. Otherwise, $C'$ should be background knowledge.

\subsection{The unplugged classroom}

\begin{figure}%[h]
  %\centering
  \includegraphics[width=0.47\textwidth]{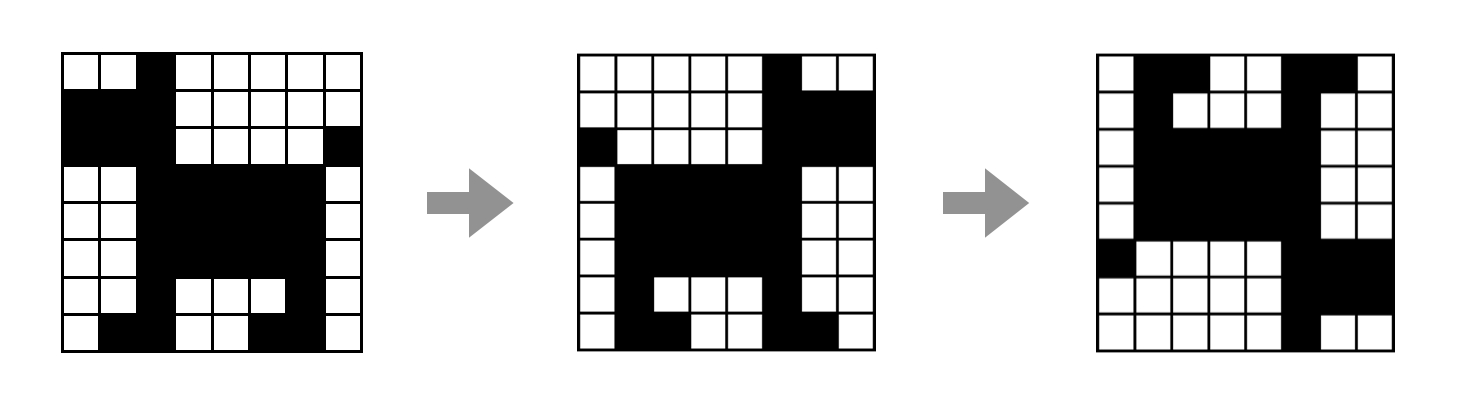}
   \caption{Bitmaps are introduced in black-and-white grids easy to draw and manipulate on the square pages of the notebook offered to \ensico\ students (Fig.~\ref{fig:leibniz}). At this early phase, manipulation is bound to bit-level operations (\emph{flip} or \emph{keep} the bit). When later on bitmaps become structured as lists, manipulation becomes list-structured \cite{Wi76}, involving operations such as (in Haskell) \emph{map}, \emph{reverse} and son on. Thus the `atomistic' start gives place to the structured view that eventually will lead to programming.\label{fig:bmap}}
  \Description{Bit maps.}
\end{figure}

As already mentioned, the adopted teaching model starts ``unplugged'' in year one, with short stories that subliminally introduce the binary system of information representation and communication (Fig.~\ref{fig:lili}). Eventually, such stories lead to black and white representation of things and animals that get \emph{transformed} into each other via black-white level operations (Fig.~\ref{fig:bmap}) using pen and paper.

Black and white will soon give room to symbols $0$ and $1$ (faster to draw than painting black squares on the notepad) that can be listed. Lists of such things (``bits'') convey the concept of a \emph{bitmap} representation, which is the first (abstract) \emph{data model} that students encounter.

With lists (finite sequences) come concepts such as order, repetition, reversing and mapping. Soon children will know about letters and numbers and lists of these come handy, giving birth to words (``strings'') and introducing the concepts of dictionary order and sorting.

\begin{figure}%[h]
  %\centering
  \includegraphics[width=0.47\textwidth]{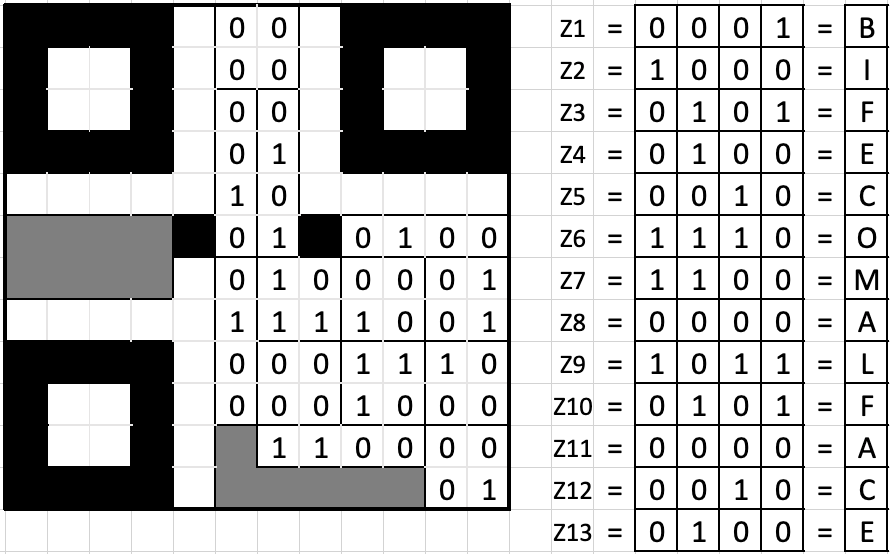}
   \caption{\ensico's toy `QR-code'. More than technology, the emphasis is on teaching \emph{concepts} \cite{Jac21}, first with pen and paper and later on the computer. This promotes digital literacy in a natural, inclusive and enjoyable way.\label{fig:qr}}
  \Description{QR-code.}
\end{figure}

Unfortunately, computers are like \lili: they cannot communicate using such letters, words and numbers, only using \emph{bits}. Representing letters by bits comes next, leading to a rich path of knowledge acquisition that, through e.g.\ "toy QR-codes" (Fig.~\ref{fig:qr}) eventually teaches message encryption and decryption --- a topic students love to do (e.g.\ C\ae sar's Cypher) on their notepads.

\begin{figure}%[h]
  %\centering
 \includegraphics[width=0.47\linewidth]{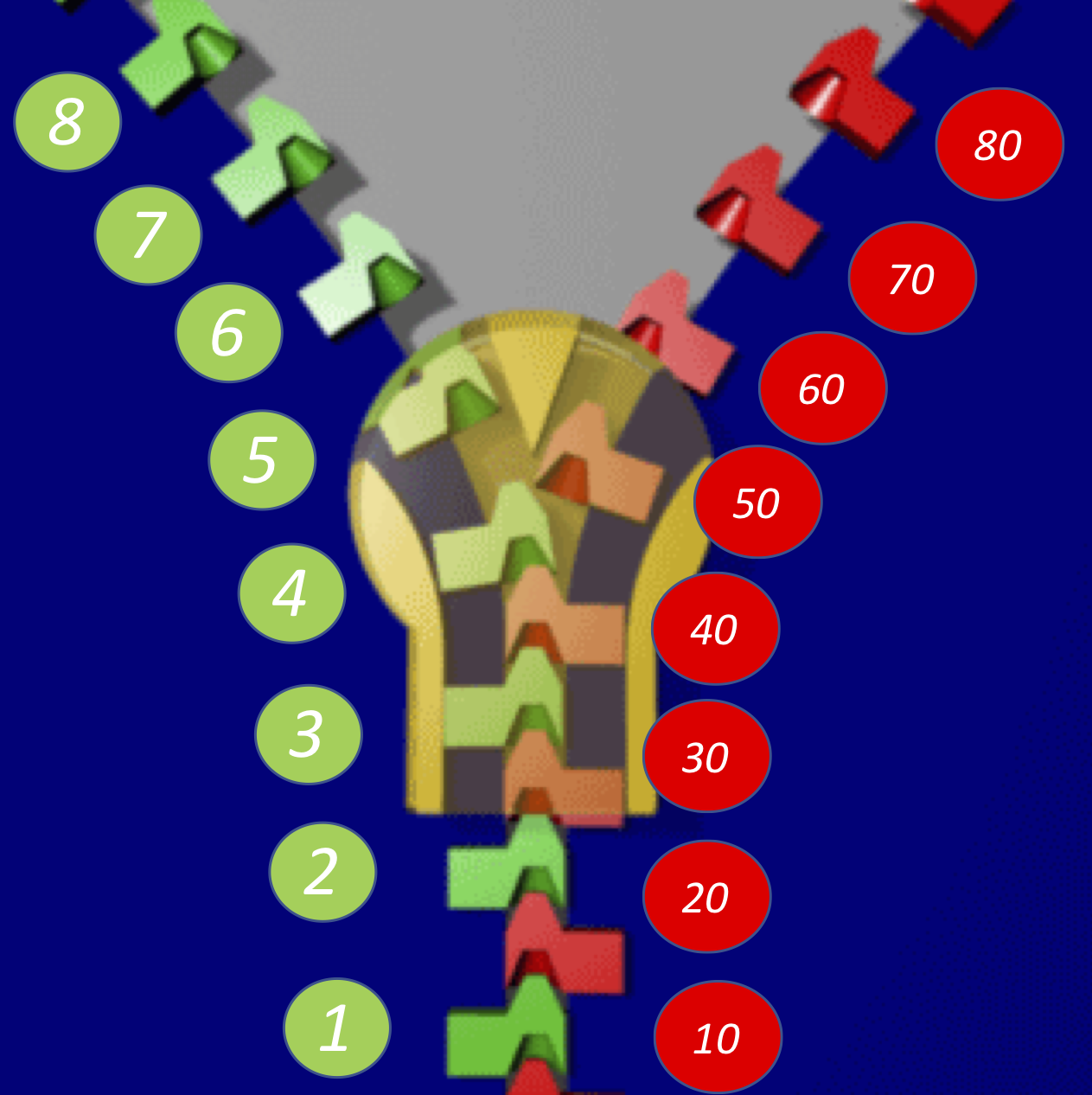}
   \caption{The ``zipper metaphor'' for introducing the concept of pairing blends nicely with the $zip$ function of the Haskell Prelude \cite{DBLP:journals/jfp/Jones03g}. \label{fig:zip}}
  \Description{zipper.}
\end{figure}

The second, basic data-structuring concept is \emph{pairing}, introduced through the ``zipper metaphor'' (Fig.~\ref{fig:zip}). Lists of pairs --- the first data-structure compound in the teaching --- are fertile ground for exploring a wealth of new concepts given by example: the ''following'' network of Instagram, roads between cities, price lists, etc. Abstract concepts so important as \emph{graphs} and \emph{key-value stores} sneakily get into the students minds just by playing with situations they know about from outside experience.

% Until the end of the school year these young students will learn how to add colors (modulo operation), how to pair numbers and colors, how to pair numeric lists (\textbf{zip} operation), how to \textbf{sort} and \textbf{reverse} lists of characters and lists of numbers and even how to keep only the first occurrence of each element of a list (\textbf{nub} operation).

% As the journey progresses into the second year, the initial binary codifications emerge in Braille, as well as the composition of \textbf{flip} and \textbf{keep} functions. And then the letters and their alphabetical order, or their numbers. And a witch who manipulates these letters and their corresponding numbers to craft cryptic messages. In other words, we have young students learning how to play with objects (or \textbf{data}), and how to make fun of them (or how to \textbf{transform} them).

\subsection{The "semi-plugged" classroom}

\begin{figure}[h]
  %\centering
  \includegraphics[width=0.5\textwidth]{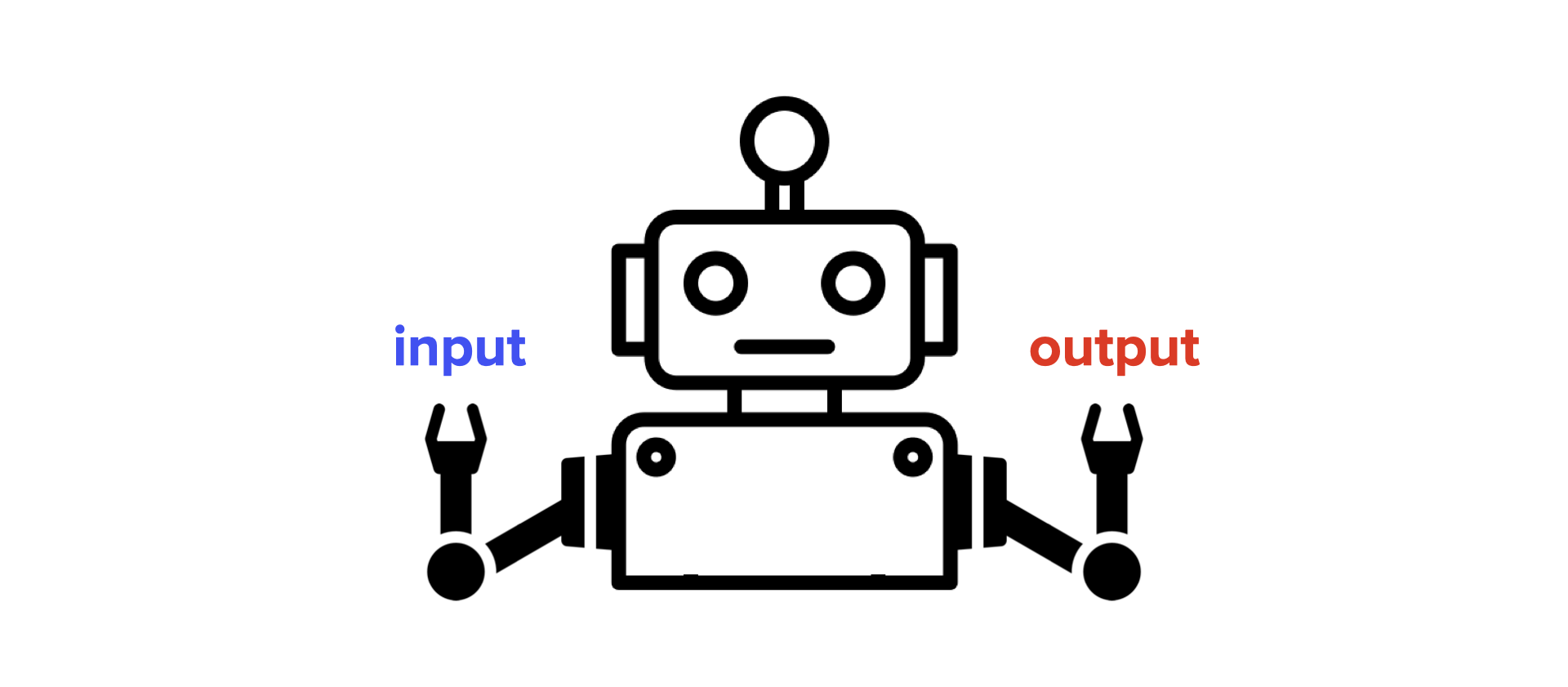}
  \caption{In the \emph{data-oriented} teaching style advocated by \ensico\ for the first steps in programming, computations are approached via data-transformers, i.e.\ ``robots'' who do the ``magic" of transforming inputs to outputs.\label{fig:robot}}
  \Description{robot.}
\end{figure}

\begin{figure*}
  %\centering
  \includegraphics[width=1.00\textwidth]{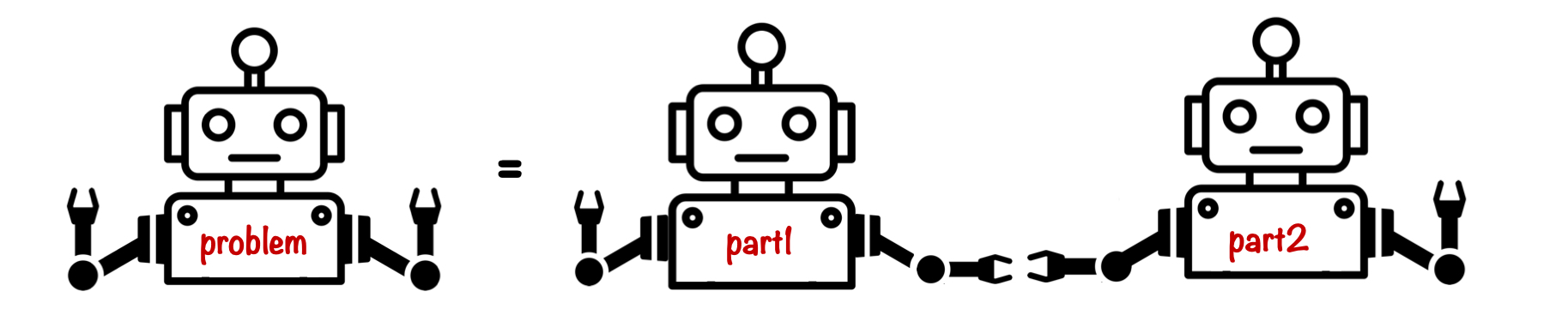}
  \caption{Robots can solve difficult problems by seeking help from others who give hands to collaborate with each other. Being able to decompose problems into sub-problems is the essence of Computational Thinking \cite{Wi06}. In functional programming this is achieved by function composition, for instance by declaring \texttt{problem = part1 >> part2} in the illustration above, where "\texttt{>>}" denotes forward (left to right) composition. This stimulates code reuse because, instead of inventing \texttt{problem} from scratch, students are invited to reuse functions \texttt{part1} and \texttt{part2} that already exist (or that they programmed before).  \label{fig:robots}}
  \Description{robots.}
\end{figure*}

Before the end of the second grade new characters are added to the narrative: \emph{robots} that have names and do the magic of transforming things into other things,
%hail from the future, bearing names encoded as numerical lists.
inputs on one hand, outputs on the other
(Figs.~\ref{fig:robot} and \ref{fig:robots}). Thus robots are a metaphor for \emph{functions}. Students engage in various activities that make them familiar with robots that now do what they were doing on paper before (e.g.\ \emph{sort}, \emph{reverse} and \emph{nub}) and a few others that do new magics, e.g.\ \emph{maximum}, \emph{sum} and \emph{length}. And, of course,  robots can join hands and dance. Then a new kind of magic comes through: by joining hands robots pass things to each other, chaining their transformative powers (Fig,~\ref{fig:robots}). In this way, students implicitly learn the essence of \emph{function composition} in the second grade.

\begin{figure}[h]
  %\centering
  \includegraphics[width=1\linewidth]{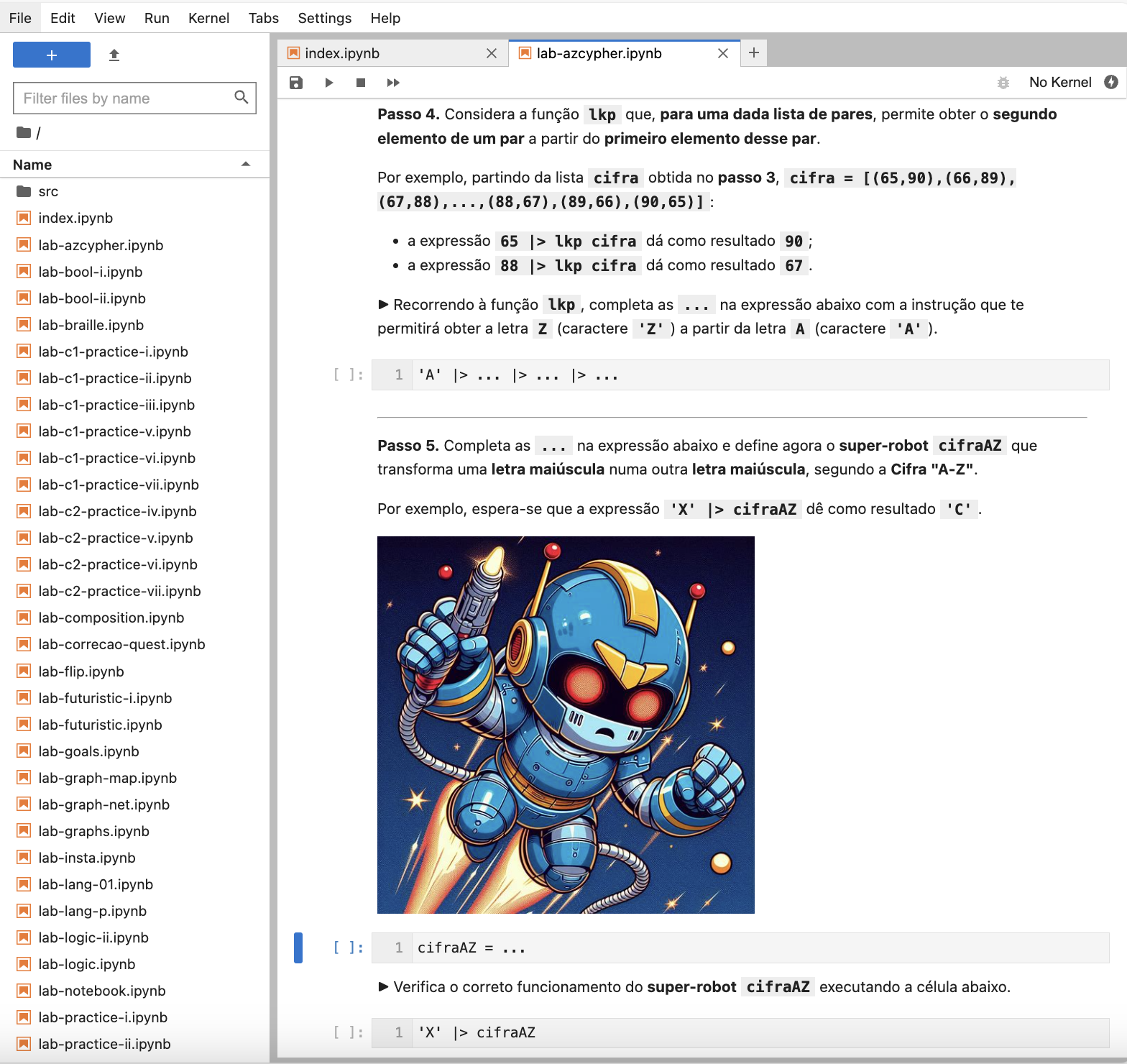}
   \caption{Robots bring a shift in the teaching: classes become \emph{semi-plugged} in the sense that students' answers on paper are validated by the teacher's computer (the only machine in the classroom). Interestingly, their reaction is often negative: \emph{I don't need a computer to sort that (short) list --- I can do it myself!} And they are right. However, when challenged with (say) sorting a list of a thousand numbers,  they will immediately agree that \jupyter\ far outperforms  them, to their amazement.
Thus they meet the concept of \emph{dematerialization}: \emph{computers are machines that help us fulfilling tasks that are too boring or complex for us to do manually}.
%(Otherwise, they are perhaps not needed...)
\label{fig:jupyter}}
  \Description{Plugged Classes}
\end{figure}

The robot metaphor is also intended to implicitly convey the shift from \emph{manual} to \emph{automatic}. Robots are machines that promise to replace the children in carrying out their assignments. How can this take place? As robots are functions in disguise, teachers can ``run'' them as \haskell\ expressions executed in the \jupyter\ platform (Fig.~\ref{fig:jupyter}).

\subsection{The plugged classroom}

\begin{figure}%[h]
  \includegraphics[width=0.95\linewidth]{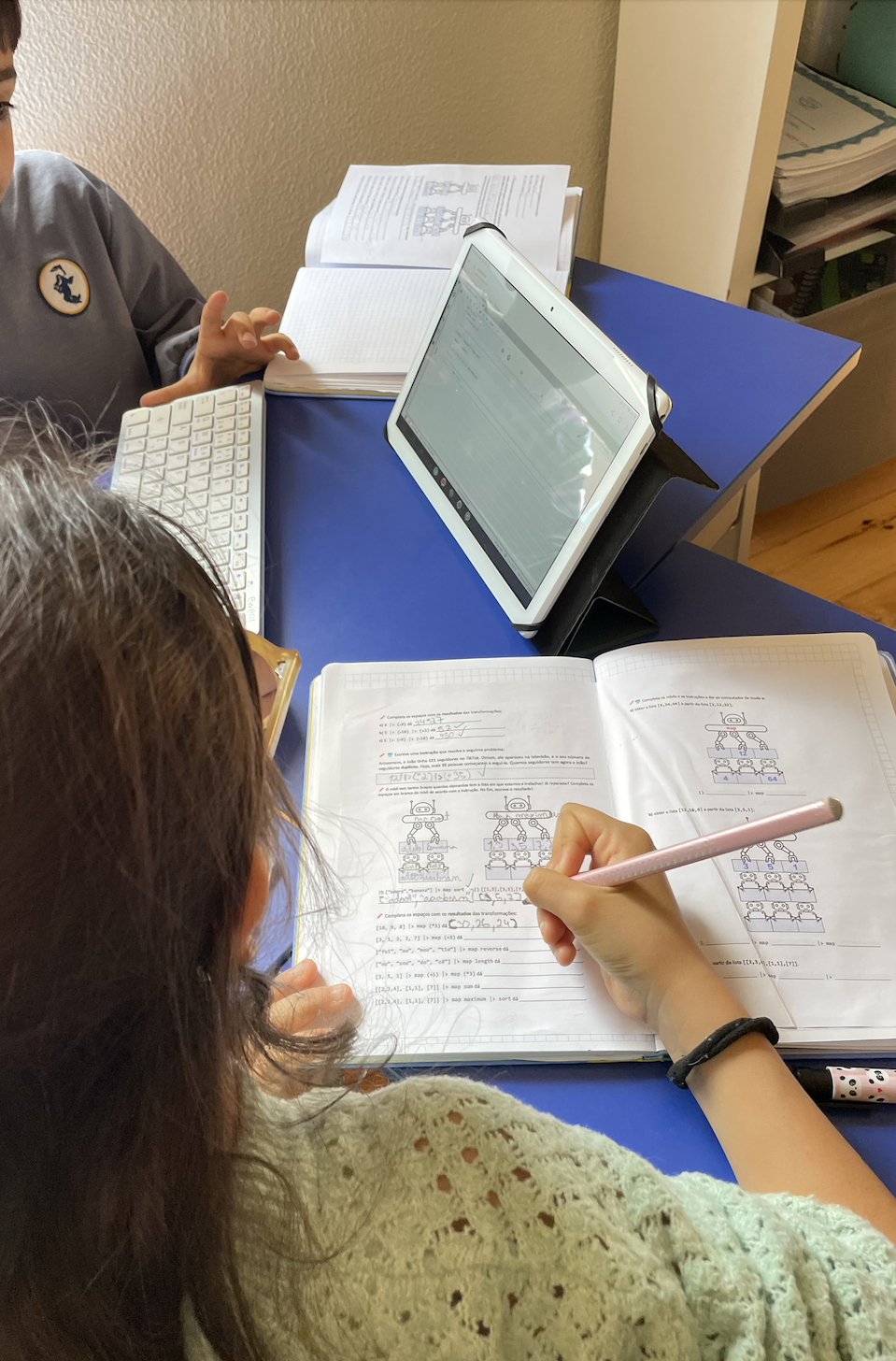}
   \caption{In plugged classes students learn by doing, solving lots of exercises available either in physical or digital notebooks. Thus "Leibniz meets Jupyter", i.e.\ the ``first computer'' of Fig.~\ref{fig:leibniz} gives room to the Jupyter platform, whose notebook cells can be interpreted and programmed. Eventually, thoughts that started expressed on paper become programs. \label{fig:plugged-classes}}
  \Description{Plugged Classes}
\end{figure}

The final modulation of the \ensico\ teaching strategy for the first two cycles is to give students direct access to the on-line \jupyter\ platform in lab sessions (Fig.~\ref{fig:jupyter} and ~\ref{fig:plugged-classes}). Towards the end of the second cycle, they become more and more familiar with the \jupyter\ environment and how to solve problems by chaining \emph{robots}, i.e.\ functions that transform data in \jupyter\ cells.

As this "plugging" them into a programming environment gradually occurs, the all-important cognitive shift from \emph{analysis to synthesis} takes place: instead of the analytical exercise of predicting the outcome of evaluating a \jupyter\ cell expression by their teacher (or doing that by themselves), students are now invited to create such expressions themselves. In other words, what might be called \emph{evaluate} mode gives room to a \emph{define} mode. This challenge is new to them and this is where the functional programming style and the syntactic simplicity of \haskell\ really helps to introduce programming (Fig.~\ref{fig:plugged-classes}).

\begin{figure}
  %\centering
  \includegraphics[width=0.47\textwidth]{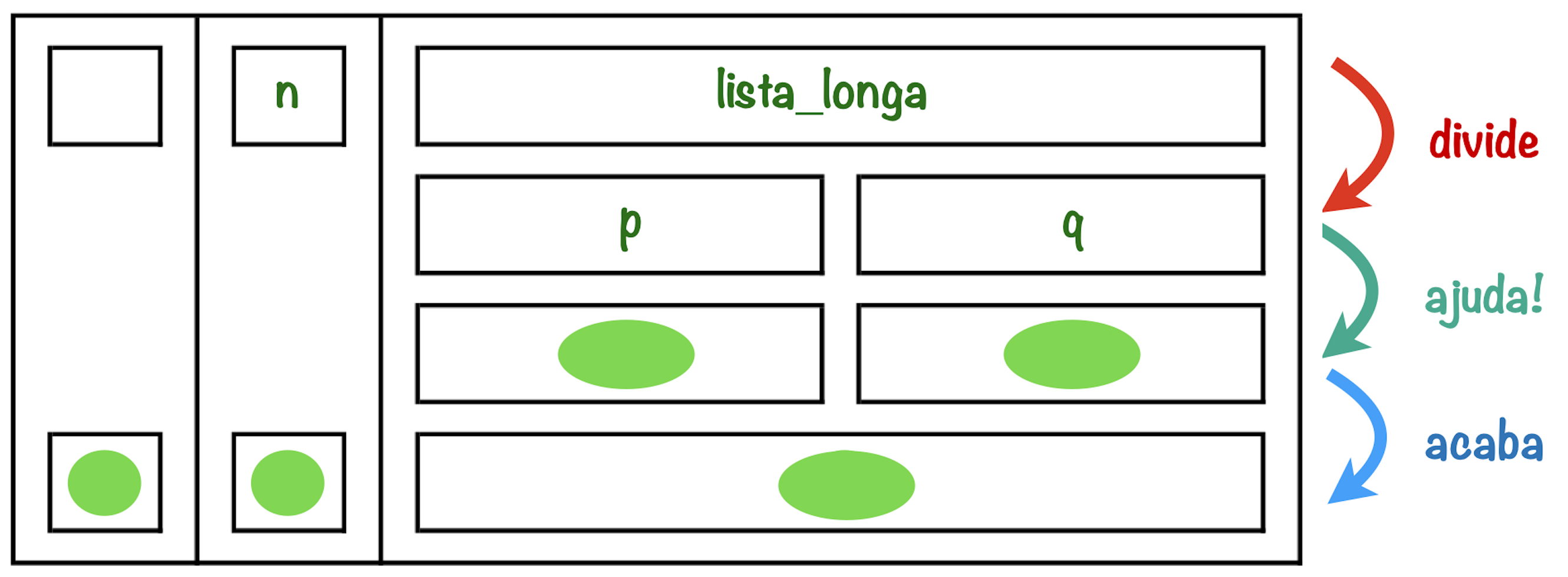}
   \caption{One of the templates used to plan (on paper) a \emph{divide and conquer} algorithm over lists late in the third cycle. The students are invited to fill in the boxes marked in green as preparation for laying down the code. The idea that one should not rush to write code but rather think about it on paper beforehand is central to the teaching, at this stage. In this way, much later at the university, students will not feel unprepared to learn the formal theories that underliy such easy-to-use templates \cite{BM97}. \label{fig:dc}}
  \Description{D&C template.}
\end{figure}

Throughout the three years of the third cycle, computer activities become more intense, enhancing analytical and problem-solving abilities through their breakdown using loops and \emph{divide and conquer} strategies (Fig.~\ref{fig:dc}) . This is also the time to diversify the technology and introduce students to other programming languages and paradigms, e.g.\ \fsharp\ and \python.\footnote{\ensico\ classroom experience in this cycle is less consolidated than that in the previous cycles because most students have not reached that level yet.}

Also significant was the creation of a first prototype of the \ensico\ on-line platform, which provided continuous access to pedagogical materials for students. Financially supported by the Belmiro de Azevedo Foundation and initially developed using \textsc{WordPress/ LearnDash}\footnote{https://www.learndash.com/}, it featured over 30 online computing classes for grades 5 and 6, along with 500 resources across six categories: slides, articles, videos, unplugged activities, \jupyter\ plugged activities and questionnaires. This platform was made accessible to 1000 students at the start of the 2022/23 academic year. Currently, it is being migrated and centralized within \textsc{Moodle}\footnote{https://moodle.org/} to ensure its availability for the final phase of the pilot (2024/25).

\section{\ensico\ and society}\label{sec:media}

To achieve its main goal \ensico\ has felt the need to contribute to explaining to the civil society why teaching computing is needed in schools. Significant efforts were made in 2021-2022 with the aim to introduce CS topics to a general audience, involving the online publication of 13 newspaper articles in collaboration with \textsc{Jornal Público} \cite{jpublico}. The total weekly views for these publications surpassed 500. The social media, e.g.\ \textsc{Instagram}\footnote{https://www.instagram.com/ensico.pt/} and \textsc{YouTube}\footnote{\url{https://www.youtube.com/channel/UC5c66xEDYKZVSgmwRk_ANow}}, are also important vehicles for the dissemination of its activities and messages to the educational community and the general public.

On March 25, 2022, \ensico\ and \textsc{APDC} organized the event \emph{"Living with Technology: Innovation in Education"} \cite{Conf22} hosted by the Faculty of Engineering of the University of Porto. Invited speakers Simon Peyton Jones and Simon Humphreys presented the \textsc{CAS UK} initiative, triggering a lively discussion on several key topics, including the significance of computational thinking, the need for computing as a compulsory topic, its influence on the scientific community, the ways in which curricula should evolve to integrate computing, and the socio-economic effects of such changes to the curricula.
%he insights, challenges, and outcomes presented by invited speakers Simon Peyton Jones and the former Executive Director, together with
The testimony of Tim Bell (from CS Unplugged) was also very important for a in-depth reflection on the dilemma of integrating computing in standard K-12 curricula. In fact, \ensico\ has been greatly influenced by these two movements and their pioneering founders.

\section{Related work}

Although CT is a topic older than 2006, it was only after Wing's paper \cite{Wi06} that it became a global trend. In 2019, Denning and Tedre wrote a book about the subject \cite{DT19} that traces it back to the 1960s and thoroughly discusses CT from various perspectives, giving useful pointers for further reading. It praises the \emph{CS Unplugged} approach \cite{DBLP:journals/cacm/Bell21} for, since the late 1990s, gaining worldwide followers and influencing the design of the ACM K-12 and code.org curricula recommendations.

Weintrop \emph{et al} \cite{Weintrop2016} argue for the inclusion of CT in mathematics and science classrooms, seeing among the main benefits for the approach the fact that it brings science and mathematics education more in line with current professional practices in these fields.

Regarding currently widespread CT as "Basic CT", or "CT 1.0", Tedre \emph{et al} \cite{TDT21} face the need for stepping further to “CT 2.0”, i.e.\ machine learning (ML) enhanced CT. They regard CT 1.0 outdated for several reasons. One of them is that it is dominated by imperative programming in e.g.\ Python, Pascal, Basic, Java, Scratch, languages that have been a “mainstay of computing education for nearly three quarters of a century”. Quantum computing and machine learning are given as examples of knowledge areas that have few counterparts in traditional computational thinking. 

The authors agree with these criticisms, which partly explain their decision to develop a \emph{functional-first} approach to CT --- a possibility that, surprisingly, seems absent from the literature. Arguing how this possibility remedies, at least in part, the criticisms made of the imperative CT mainstay is, however, outside the scope of this report and will have to wait for another time.

\section{Summary and Future Work}\label{sec:eval}
Since its very start in 2020, the \ensico\ pilot project has been a ``hands-on'', out-of-the-box experiment. Portugal is behind in computing education and time cannot be wasted.
%Framed into the computing education trend %\fbox{
%    VER %\cite{\cite{DBLP:journals/eait/WebbDBKRCS1%7,DBLP:journals/jcal/SunHZ21,Xu2023,VF20}}
%    },
%it may be regarded as an out-of-the-box %project, \emph{learning by doing} and not %just going \emph{by the book}.
Its unplugged flavour fights the ``digital drug" syndrome that is afflicting so many schools and social environments. The worst thing computing education could do is to \emph{spoil the child's brain} --- it should rather contribute to its development and creativity \cite{Kr07}. 

\begin{figure}%[h]
  %\centering
  \includegraphics[width=1\linewidth]{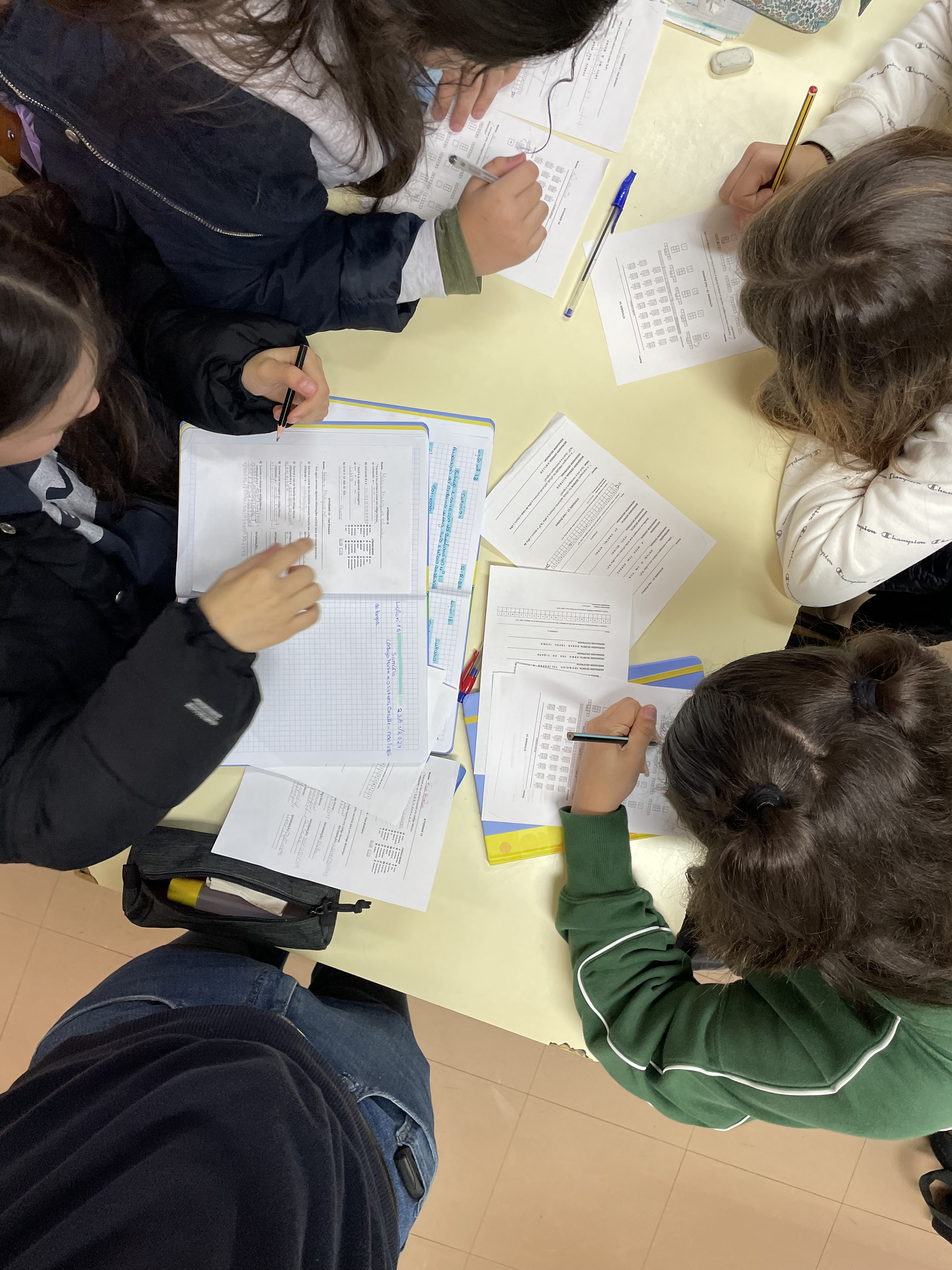}
  \caption{Computing is far more than using computers and can be taught without them. This view is inclusive in the sense that even the least equipped schools can implement \ensico's teaching. Above all, it promotes authentic digital literacy and not just technology skills. In this respect, it is also gender inclusive as it does not assume the technological proclivity that boys often have when compared to girls at a young age.\label{fig:girls}}
  \Description{All inclusive computing classes}
\end{figure}

Above all, we believe that \emph{learning should be fun}, not only for students but also for teachers, who in this way get motivated and engaged in the project ideas. Furthermore, students and parents should not be fooled by the idea that computing is a form of entertainment at school - studying computing is “hard work”, just like mathematics, arts and science.

Since 2020 \ensico\ has collected many data concerning its pilot project but thoroughly analysing them is out of the scope of this experience report. Some plain figures can nevertheless be given, further to those of Table \ref{tab:pilot}. In the current academic year, 2000+ slides
%
% card . discollect . map (id >< parse) . drop 2 . prj fourth (!!4) $ xlsx
%
are being used in classes altogether (one hour per week). Many of these slides are targeted to teachers only, containing notes, explanations and suggestions about the slides to display in class.\footnote{The prospect of converting all this material in textbooks and exercise books is currently under consideration. Also, a formal curriulum for Computing is currently being prepared according to the current standards \cite{vandenAkker2003}.}
This teaching material is supported by more than 100 \jupyter\ notebooks, 150 unplugged activities and 150 evaluation sheets.

\begin{figure}%[h]
  \centering
  \includegraphics[width=1.00\linewidth]{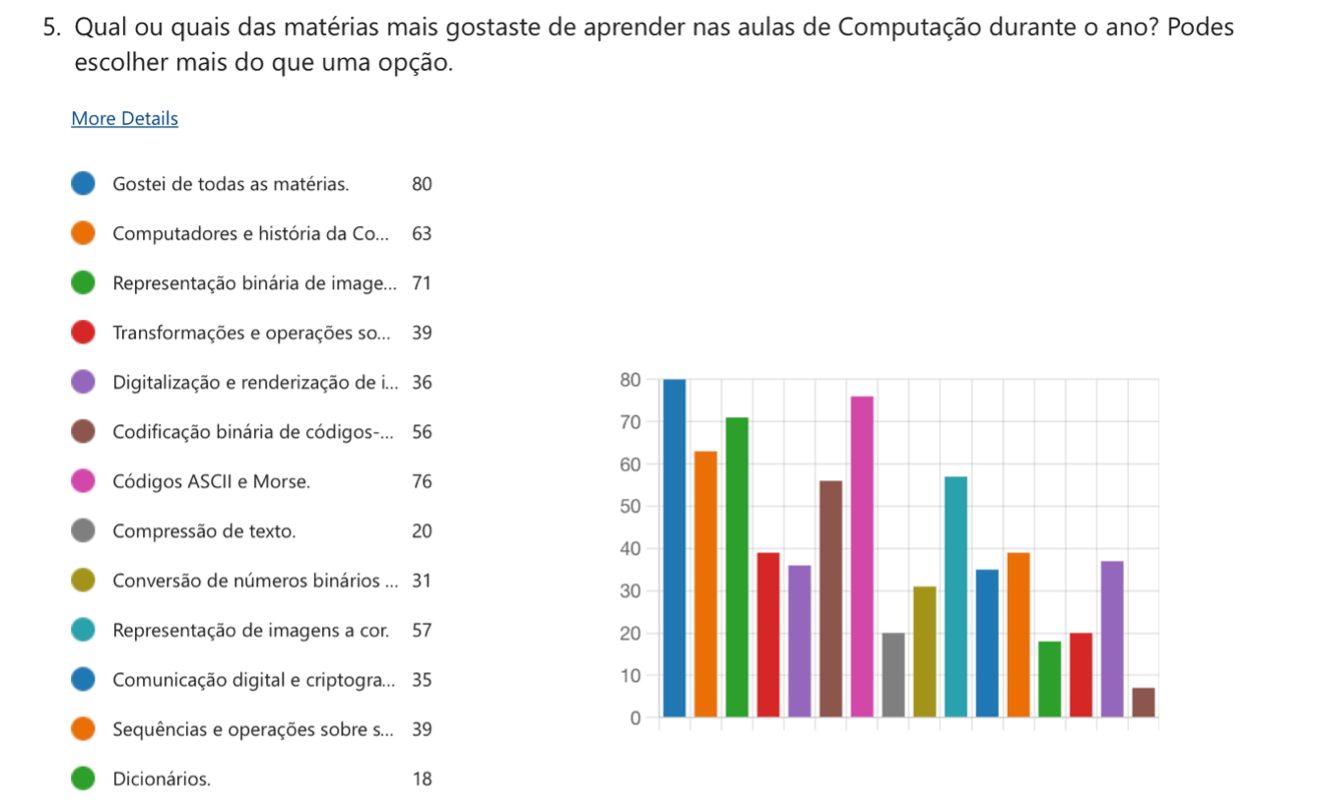}
  \caption{Sample questionnaire enquiring students about which subjects they enjoyed most (2021/22). Most students liked the course as a whole. Topics such as ASCII, Morse code and binary representation of images were the most favored.\label{fig:questionnaire}}
  \Description{questionnaire}
\end{figure}

Activity reports are sent to \ensico\ sponsors every school year. These reports typically include multi-dimensional statistics and opinion polls for students, teachers and parents (Fig.\ref{fig:questionnaire}).

For the next four years, \ensico\ will conduct a rigorous impact assessment in collaboration with {Nova School of Business and Economics}\footnote{https://www.novasbe.unl.pt/en/} in order to evaluate the influence on students of having computing classes from grade 7 (3rd cycle of basic school) to grade 10 (1st year of secondary school). The outcome of such a joint venture is regarded as a key factor for the national-wide implementation of \ensico's program in Portugal.

Preparations are underway for the final school year of the pilot program (2024/25). Over ten thousand students are expected to participate in computing classes, some of them reaching their fourth year of computing. Moreover, if everything goes according to the plan, such students will be the first in Portugal to finish mandatory schooling with a complete, K-12 education in computer science.
%knowledge is now in the third grade and at the beginning of their third year of computing classes.

In the short term, two significant milestones are to be pursued. The first and foremost is the government's decision on the adoption of \ensico's program into the national basic and secondary school curricula. The second, equally ambitious, concerns the international adoption of our program. Observing the various initiatives across Europe, we are confident that many could benefit from our experience in Portugal.

Last but not least, the prospect of a nation-wide K-12 education on computing will bring new opportunities for the teaching of more ambitious subjects at higher education levels. In other words, it has the potential to revolutionize and update the current IT pedagogical content taught in universities and polytechnics.
\ensico\ would like to be involved in discussions in this regard, contributing to generating a new dynamic with valuable economic potential in Portugal and throughout Europe.

\section{Acknowledgments}\label{sec:ackn}

\ensico\ has been counting on a growing range of partners, among which we highlight .PT (management entity of Portugal's top-level domain on the Internet and co-founder of \ensico) and the High Patronage of the President of the Portuguese Republic. \ensico\ also has had the support of the municipalities of Oporto and Cascais and of companies, foundations, R\&D organizations and universities such as SONAE SGPS, NOS SGPS, Calouste Gulbenkian Foundation, Belmiro de Azevedo Foundation and University of Porto, among others. Scientific collaboration and advice by INESC TEC and the University of Minho is also gratefully acknowledged. Special thanks are due to Minho's CS department for hosting and running the Jupyter infra-structure that supports all \ensico\ classes.

Very special thanks go to the founders and all collaborators of \ensico, in particular the extraordinary group of 'Master Teachers' who have been teaching the students attending their initial computing classes, as well as the teachers conducting their first computing lessons. Also, we would like to express our gratitude to Andrew Smith and Martina Mayrhofer for their unwavering commitment, which was instrumental in the successful completion of the ERASMUS+ project \textsc{CS4All}, despite the extremely tight schedule.

We also want to record our appreciation and immense gratitude to
Pedro Guedes de Oliveira (University of Porto Emeritus Professor),
Luísa Ribeiro Lopes (Chair of the Board of Directors of DNS.pt),
Ricardo Valente (Porto City Councillor for Economy, Employment and Entrepreneurship),
João Gunther Amaral (Executive Board Member at SONAE) and
Miguel Pinto Luz
(Minister of Infrastructure and Housing, former Cascais Councillor).
Without their contribution and support the \ensico\ project could not have been realized and achieved its current prestigious status.

Our deepest thanks go to our students, parents and guardians, teachers, and the school boards who have welcomed us warmly and professionally. Their good will and trust have actively contributed to the realization of the \ensico\ mission.

J.N. Oliveira thanks Roland Backhouse for introducing him to the CS Unplugged initiative \cite{DBLP:journals/cacm/Bell21} that has so much influenced the approach put forward in this report.

Finally, the authors thank the WIPSCE'24 reviewers for their comments and criticisms that helped improve an earlier version of this report.
 
%\begin{acks}
%To Robert, for the bagels and explaining CMYK and color spaces.
%\end{acks}

\bibliographystyle{ACM-Reference-Format}
%\bibliography{bib}


\begin{thebibliography}{27}

%%% the Web address.

\ifx \showCODEN    \undefined \def \showCODEN     #1{\unskip}     \fi
\ifx \showDOI      \undefined \def \showDOI       #1{#1}\fi
\ifx \showISBNx    \undefined \def \showISBNx     #1{\unskip}     \fi
\ifx \showISBNxiii \undefined \def \showISBNxiii  #1{\unskip}     \fi
\ifx \showISSN     \undefined \def \showISSN      #1{\unskip}     \fi
\ifx \showLCCN     \undefined \def \showLCCN      #1{\unskip}     \fi
\ifx \shownote     \undefined \def \shownote      #1{#1}          \fi
\ifx \showarticletitle \undefined \def \showarticletitle #1{#1}   \fi
\ifx \showURL      \undefined \def \showURL       {\relax}        \fi
% invisible to TeX
\providecommand\bibfield[2]{#2}
\providecommand\bibinfo[2]{#2}
\providecommand\natexlab[1]{#1}
\providecommand\showeprint[2][]{arXiv:#2}

\bibitem[Bell(2021)]%
        {DBLP:journals/cacm/Bell21}
\bibfield{author}{\bibinfo{person}{Tim Bell}.} \bibinfo{year}{2021}\natexlab{}.
\newblock \showarticletitle{{CS} unplugged or coding classes?}
\newblock \bibinfo{journal}{\emph{Commun. {ACM}}} \bibinfo{volume}{64},
  \bibinfo{number}{5} (\bibinfo{year}{2021}), \bibinfo{pages}{25--27}.
\newblock
\urldef\tempurl%
\url{https://doi.org/10.1145/3457195}
\showDOI{\tempurl}


\bibitem[Bird and de~Moor(1997)]%
        {BM97}
\bibfield{author}{\bibinfo{person}{R. Bird} {and} \bibinfo{person}{O. de
  Moor}.} \bibinfo{year}{1997}\natexlab{}.
\newblock \bibinfo{booktitle}{\emph{Algebra of Programming}}.
\newblock \bibinfo{publisher}{Prentice-Hall}.
\newblock
\newblock
\shownote{ISBN: 978-0-13-507245-5}.


\bibitem[Bird and Wadler(1988)]%
        {BW88}
\bibfield{author}{\bibinfo{person}{R. Bird} {and} \bibinfo{person}{P. Wadler}.}
  \bibinfo{year}{1988}\natexlab{}.
\newblock \bibinfo{booktitle}{\emph{Introduction to Functional Programming}}.
\newblock \bibinfo{publisher}{Prentice-Hall}.
\newblock


\bibitem[Bugliarello(2003)]%
        {Bu03}
\bibfield{author}{\bibinfo{person}{G. Bugliarello}.}
  \bibinfo{year}{2003}\natexlab{}.
\newblock \showarticletitle{A New Trivium and Quadrivium}.
\newblock \bibinfo{journal}{\emph{Bulletin of Science, Technology \& Society}}
  \bibinfo{volume}{23}, \bibinfo{number}{2} (\bibinfo{year}{2003}),
  \bibinfo{pages}{106--113}.
\newblock
\urldef\tempurl%
\url{https://doi.org/10.1177/0270467603251296}
\showURL{%
\tempurl}


\bibitem[Denning and Tedre(2019)]%
        {DT19}
\bibfield{author}{\bibinfo{person}{P.J. Denning} {and} \bibinfo{person}{M.
  Tedre}.} \bibinfo{year}{2019}\natexlab{}.
\newblock \bibinfo{booktitle}{\emph{Computational Thinking}}.
\newblock \bibinfo{publisher}{{MIT} Press}.
\newblock
\showISBNx{9780262536561}
\urldef\tempurl%
\url{https://mitpress.mit.edu/9780262536561/computational-thinking/}
\showURL{%
\tempurl}


\bibitem[\ensico(2022)]%
        {jpublico}
\bibfield{author}{\bibinfo{person}{\ensico}.}
  \bibinfo{year}{2021-2022}\natexlab{}.
\newblock \bibinfo{title}{Público na Escola}.
\newblock
  \bibinfo{howpublished}{\url{https://www.publico.pt/publico-na-escola/grafos}}.
\newblock
\newblock
\shownote{Accessed: \today}.


\bibitem[ENSICO and APDC(2022)]%
        {Conf22}
\bibfield{author}{\bibinfo{person}{ENSICO} {and} \bibinfo{person}{APDC}.}
  \bibinfo{year}{2022}\natexlab{}.
\newblock \bibinfo{title}{Living with Technology: Innovation in Education}.
\newblock \bibinfo{howpublished}{URL:
  \url{https://www.apdc.pt/iniciativas/agenda-apdc/conference--living-with-technology-innovation-in-education}}.
\newblock
\newblock
\shownote{Accessed: \today}.


\bibitem[\ensico\ Association(2024)]%
        {Ensico}
\bibfield{author}{\bibinfo{person}{The \ensico\ Association}.}
  \bibinfo{year}{2024}\natexlab{}.
\newblock \bibinfo{title}{Home page}.
\newblock \bibinfo{howpublished}{URL: \url{https://ensico.pt/}}.
\newblock
\newblock
\shownote{Accessed: \today}.


\bibitem[for All(2023)]%
        {CS4ALL}
\bibfield{author}{\bibinfo{person}{Computer~Science for All}.}
  \bibinfo{year}{2023}\natexlab{}.
\newblock \bibinfo{title}{Home page}.
\newblock \bibinfo{howpublished}{URL: \url{http://cs4all.pt/}}.
\newblock
\newblock
\shownote{Accessed: \today}.


\bibitem[for All(2022)]%
        {I4A}
\bibfield{author}{\bibinfo{person}{Informatics for All}.}
  \bibinfo{year}{2022}\natexlab{}.
\newblock \bibinfo{title}{The Informatics Reference Framework for School}.
\newblock \bibinfo{howpublished}{URL:
  \url{https://www.informaticsforall.org/the-informatics-reference-framework-for-school-release-february-2022/}}.
\newblock
\newblock
\shownote{Accessed: \today}.


\bibitem[Gibiansky(2007)]%
        {Gi07}
\bibfield{author}{\bibinfo{person}{A. Gibiansky}.}
  \bibinfo{year}{2007}\natexlab{}.
\newblock \bibinfo{title}{ihaskell: A Haskell backend kernel for the Jupyter
  project}.
\newblock \bibinfo{howpublished}{Hackage -
  \url{https://hackage.haskell.org/package/ihaskell}}.
\newblock
\newblock
\shownote{Accessed: \today}.


\bibitem[Jackson(2021)]%
        {Jac21}
\bibfield{author}{\bibinfo{person}{D. Jackson}.}
  \bibinfo{year}{2021}\natexlab{}.
\newblock \bibinfo{booktitle}{\emph{The Essence of Software: Why Concepts
  Matter for Great Design}}.
\newblock \bibinfo{publisher}{Princeton University Press}.
\newblock


\bibitem[Jones(2003)]%
        {DBLP:journals/jfp/Jones03g}
\bibfield{author}{\bibinfo{person}{S.L.~Peyton Jones}.}
  \bibinfo{year}{2003}\natexlab{}.
\newblock \showarticletitle{Haskell 98: Standard Prelude}.
\newblock \bibinfo{journal}{\emph{J. Funct. Program.}} \bibinfo{volume}{13},
  \bibinfo{number}{1} (\bibinfo{year}{2003}), \bibinfo{pages}{103--124}.
\newblock
\urldef\tempurl%
\url{https://doi.org/10.1017/S0956796803001011}
\showDOI{\tempurl}


\bibitem[Kolling et~al\mbox{.}(2013)]%
        {kolling2013bringing}
\bibfield{author}{\bibinfo{person}{M. Kolling}, \bibinfo{person}{T. Crick},
  \bibinfo{person}{S. Peyton~Jones}, \bibinfo{person}{S. Humphreys}, {and}
  \bibinfo{person}{S. Sentance}.} \bibinfo{year}{2013}\natexlab{}.
\newblock \showarticletitle{Bringing computer science back into schools:
  lessons from the {UK}}. In \bibinfo{booktitle}{\emph{The 44th {ACM} Technical
  Symposium on Computer Science Education, {SIGCSE} 2013, Denver, CO, USA,
  March 6-9, 2013} (\bibinfo{edition}{sigcse’13} ed.)}.
  \bibinfo{publisher}{{ACM}}, \bibinfo{pages}{269--274}.
\newblock
\urldef\tempurl%
\url{https://doi.org/10.1145/2445196.2445277}
\showDOI{\tempurl}


\bibitem[Kramer(2007)]%
        {Kr07}
\bibfield{author}{\bibinfo{person}{J. Kramer}.}
  \bibinfo{year}{2007}\natexlab{}.
\newblock \showarticletitle{Is abstraction the key to computing?}
\newblock \bibinfo{journal}{\emph{Commun. {ACM}}} \bibinfo{volume}{50},
  \bibinfo{number}{4} (\bibinfo{year}{2007}), \bibinfo{pages}{36--42}.
\newblock
\urldef\tempurl%
\url{https://doi.org/10.1145/1232743.1232745}
\showDOI{\tempurl}


\bibitem[Lakoff and Johnson(1980)]%
        {LJ80}
\bibfield{author}{\bibinfo{person}{G. Lakoff} {and} \bibinfo{person}{M.
  Johnson}.} \bibinfo{year}{1980}\natexlab{}.
\newblock \bibinfo{booktitle}{\emph{Metaphors we live by}}.
\newblock \bibinfo{publisher}{University of Chicago Press},
  \bibinfo{address}{Chicago}.
\newblock
\showISBNx{978-0-226-46800-6}


\bibitem[MarketSplash(2024)]%
        {MarketSplash24}
\bibfield{author}{\bibinfo{person}{MarketSplash}.}
  \bibinfo{year}{2024}\natexlab{}.
\newblock \bibinfo{title}{F\# vs Haskell: Uncovering Their Unique Language
  Traits And Use Cases}.
\newblock
\newblock
\urldef\tempurl%
\url{https://marketsplash.com/f-vs-haskell/}
\showURL{%
Retrieved \today from \tempurl}


\bibitem[on~Computing~Curricula(2001)]%
        {ACM01}
\bibfield{author}{\bibinfo{person}{The Joint {ACM/IEEE}-CS Task~Force on
  Computing~Curricula}.} \bibinfo{year}{2001}\natexlab{}.
\newblock \bibinfo{booktitle}{\emph{Computing Curricula 2001: Computer Science
  --- Final Report}}.
\newblock \bibinfo{type}{{T}echnical {R}eport}.
  \bibinfo{institution}{Association for Computing Machinery and IEEE Computer
  Society}.
\newblock


\bibitem[Sun et~al\mbox{.}(2021)]%
        {DBLP:journals/jcal/SunHZ21}
\bibfield{author}{\bibinfo{person}{L. Sun}, \bibinfo{person}{L. Hu}, {and}
  \bibinfo{person}{D. Zhou}.} \bibinfo{year}{2021}\natexlab{}.
\newblock \showarticletitle{Which way of design programming activities is more
  effective to promote {K-12} students' computational thinking skills? {A}
  meta-analysis}.
\newblock \bibinfo{journal}{\emph{J. Comput. Assist. Learn.}}
  \bibinfo{volume}{37}, \bibinfo{number}{4} (\bibinfo{year}{2021}),
  \bibinfo{pages}{1048--1062}.
\newblock
\urldef\tempurl%
\url{https://doi.org/10.1111/JCAL.12545}
\showDOI{\tempurl}


\bibitem[Tedre et~al\mbox{.}(2021)]%
        {TDT21}
\bibfield{author}{\bibinfo{person}{M. Tedre}, \bibinfo{person}{P.J. Denning},
  {and} \bibinfo{person}{T. Toivonen}.} \bibinfo{year}{2021}\natexlab{}.
\newblock \showarticletitle{{CT} 2.0}. In \bibinfo{booktitle}{\emph{Koli
  Calling '21: 21st Koli Calling International Conference on Computing
  Education Research, Joensuu, Finland, November 18 - 21, 2021}},
  \bibfield{editor}{\bibinfo{person}{Otto Sepp{\"{a}}l{\"{a}}} {and}
  \bibinfo{person}{Andrew Petersen}} (Eds.). \bibinfo{publisher}{{ACM}},
  \bibinfo{pages}{3:1--3:8}.
\newblock
\urldef\tempurl%
\url{https://doi.org/10.1145/3488042.3488053}
\showDOI{\tempurl}


\bibitem[van~den Akker(2003)]%
        {vandenAkker2003}
\bibfield{author}{\bibinfo{person}{J. van~den Akker}.}
  \bibinfo{year}{2003}\natexlab{}.
\newblock \bibinfo{booktitle}{\emph{Curriculum Perspectives: An Introduction}}.
\newblock \bibinfo{publisher}{Springer Netherlands},
  \bibinfo{address}{Dordrecht}, \bibinfo{pages}{1--10}.
\newblock
\showISBNx{978-94-017-1205-7}
\urldef\tempurl%
\url{https://doi.org/10.1007/978-94-017-1205-7_1}
\showDOI{\tempurl}


\bibitem[Vegas and Fowler(2020)]%
        {VF20}
\bibfield{author}{\bibinfo{person}{E. Vegas} {and} \bibinfo{person}{B.
  Fowler}.} \bibinfo{year}{2020}\natexlab{}.
\newblock \bibinfo{title}{What do we know about the expansion of K-12 computer
  science education? A review of the evidence}.
\newblock \bibinfo{howpublished}{The The Brookings Institution
  (\url{https://www.brookings.edu})\footnote{\url{https://www.brookings.edu/articles/what-do-we-know-about-the-expansion-of-k-12-computer-science-education}.}}.
\newblock
\newblock
\shownote{Accessed: \today}.


\bibitem[Webb et~al\mbox{.}(2017)]%
        {DBLP:journals/eait/WebbDBKRCS17}
\bibfield{author}{\bibinfo{person}{M. Webb}, \bibinfo{person}{N. Davis},
  \bibinfo{person}{T. Bell}, \bibinfo{person}{Y.J. Katz}, \bibinfo{person}{N.
  Reynolds}, \bibinfo{person}{D.P. Chambers}, {and} \bibinfo{person}{M.M.
  Syslo}.} \bibinfo{year}{2017}\natexlab{}.
\newblock \showarticletitle{Computer science in {K-12} school curricula of the
  2lst century: Why, what and when?}
\newblock \bibinfo{journal}{\emph{Educ. Inf. Technol.}} \bibinfo{volume}{22},
  \bibinfo{number}{2} (\bibinfo{year}{2017}), \bibinfo{pages}{445--468}.
\newblock
\urldef\tempurl%
\url{https://doi.org/10.1007/S10639-016-9493-X}
\showDOI{\tempurl}


\bibitem[Weintrop et~al\mbox{.}(2016)]%
        {Weintrop2016}
\bibfield{author}{\bibinfo{person}{D. Weintrop}, \bibinfo{person}{E. Beheshti},
  \bibinfo{person}{M. Horn}, \bibinfo{person}{K. Orton}, \bibinfo{person}{K.
  Jona}, \bibinfo{person}{L. Trouille}, {and} \bibinfo{person}{U. Wilensky}.}
  \bibinfo{year}{2016}\natexlab{}.
\newblock \showarticletitle{Defining Computational Thinking for Mathematics and
  Science Classrooms}.
\newblock \bibinfo{journal}{\emph{Journal of Science Education and Technology}}
  \bibinfo{volume}{25}, \bibinfo{number}{1} (\bibinfo{date}{01 Feb}
  \bibinfo{year}{2016}), \bibinfo{pages}{127--147}.
\newblock


\bibitem[Wing(2006)]%
        {Wi06}
\bibfield{author}{\bibinfo{person}{J.M. Wing}.}
  \bibinfo{year}{2006}\natexlab{}.
\newblock \showarticletitle{Computational thinking}.
\newblock \bibinfo{journal}{\emph{Commun. {ACM}}} \bibinfo{volume}{49},
  \bibinfo{number}{3} (\bibinfo{year}{2006}), \bibinfo{pages}{33--35}.
\newblock
\urldef\tempurl%
\url{https://doi.org/10.1145/1118178.1118215}
\showDOI{\tempurl}


\bibitem[Wirth(1976)]%
        {Wi76}
\bibfield{author}{\bibinfo{person}{N. Wirth}.} \bibinfo{year}{1976}\natexlab{}.
\newblock \bibinfo{booktitle}{\emph{Algorithms + Data Structures = Programs}}.
\newblock \bibinfo{publisher}{Prentice-Hall}.
\newblock


\bibitem[Xu et~al\mbox{.}(2023)]%
        {Xu2023}
\bibfield{author}{\bibinfo{person}{Enwei Xu}, \bibinfo{person}{Wei Wang}, {and}
  \bibinfo{person}{Qingxia Wang}.} \bibinfo{year}{2023}\natexlab{}.
\newblock \showarticletitle{A meta-analysis of the effectiveness of programming
  teaching in promoting K-12 students' computational thinking}.
\newblock \bibinfo{journal}{\emph{Education and Information Technologies}}
  \bibinfo{volume}{28}, \bibinfo{number}{6} (\bibinfo{date}{01 Jun}
  \bibinfo{year}{2023}), \bibinfo{pages}{6619--6644}.
\newblock
\urldef\tempurl%
\url{https://doi.org/10.1007/s10639-022-11445-2}
\showDOI{\tempurl}


\end{thebibliography}


\end{document}